\documentclass[aps,pre,reprint, superscriptaddress]{revtex4-1}

\usepackage{graphicx}
\usepackage{amsmath}
\usepackage{amsfonts}
\usepackage{amssymb}
\usepackage{bm}
\usepackage{color}
\usepackage{soul} 
\usepackage{amssymb}
\usepackage{wasysym}
\usepackage{dsfont}
\usepackage{multirow}
\usepackage{epsfig}
\usepackage{bm}
\usepackage{amsmath,amsthm,amssymb}
\usepackage{array,booktabs,longtable}
\usepackage{mathrsfs}
\usepackage{subfigure}
\usepackage{color}
\usepackage{tikz}
\usetikzlibrary{decorations.markings}
\usetikzlibrary{calc,shapes,positioning,arrows,snakes}
\usepackage{array}
\usepackage{verbatim}
\usepackage{dsfont}
\usepackage{ifthen}

\newcommand{\red}{\textcolor{black}}

\newcommand{\rmn}{{\rm n}}
\newcommand{\rmt}{{\rm t}}

\begin{document}

\title{\red{Numerical} simulation of universal morphogenesis of fluid interface deformations\\driven by radiation pressure}


\author{Hugo Chesneau}
\author{Hamza Chraibi}
\author{Nicolas Bertin}
\author{Julien Petit}
\author{Jean-Pierre Delville}
\author{Etienne Brasselet}
\email{etienne.brasselet@u-bordeaux.fr}
\affiliation{Univ. Bordeaux, CNRS, LOMA, UMR 5798, F-33400 Talence, France}
\author{R\'egis Wunenburger$^{\, 1, \,}$}
\email{regis.wunenburger@sorbonne-universite.fr}
\affiliation{Sorbonne Universit\'e, CNRS, Institut Jean Le Rond d'Alembert, F-75005 Paris, France
}

\begin{abstract}
We report on \red{numerical} simulation of fluid interface deformations induced either by acoustic or optical radiation pressure. This is done by solving simultaneously the scalar wave propagation equation and the two-phase flow equations using the boundary element method. Using dimensional analysis, we show that interface deformation morphogenesis is universal, i.e. depends on the same dimensionless parameters in acoustics and electromagnetics. We numerically investigate a few selected phenomena---in particular the shape of large deformations, the slenderness transition and its hysteresis--- and compare with existing and novel experimental observations. Qualitative agreement between the numerical simulations and experiments is found when the mutual interaction between wave propagation and wave-induced deformations is taken into account. Our results demonstrate the leading role of the radiation pressure in morphogenesis of fluid interface deformations and the importance of the propagation-deformation interplay .
\end{abstract}

%
\maketitle

\section{Introduction}
In 1939, Hertz and Mende showed that fluid interfaces were deformed by intense ultrasonic beams~\cite{HertzandMende}. In 1973, Ashkin and Dziedzic~\cite{Ashkin1973} demonstrated the same phenomenon using focused optical beams. Using extremely soft fluid interfaces, Casner and Delville~\cite{Casner2001} observed `large' interface deformations (i.e., whose height is comparable to or larger than the beam diameter) using optical beams. Since then, various kinds of interface deformations have been observed using either acoustic or optical beams: bell-shaped deformations~\cite{Casner2001,Issenmann2008}, stepped deformations~\cite{casner2003b,Issenmann2008,Bertin2012}, needle-like deformations~\cite{Casner2003,Issenmann2008,wunenburger2011}, liquid bridges~\cite{Casner2004,Brasselet2008,Brasselet2008_2,Bertin2010} and cones~\cite{girot2019}. Since Maxwell's and Lord Rayleigh's pioneering works
~\cite{maxwell1873treatise,strutt1902}, it is well established that the main mechanical effect of an acoustic (AC) or electromagnetic (EM) wave on a liquid-liquid interface is a normal stress called radiation pressure~\cite{Brevik1979,borgnis1953}. Since then, several other mechanical effects of acoustic and electromagnetic waves on liquids and liquid interfaces have been identified, in particular in presence of wave absorption or wave scattering by liquids. Namely, one can mention situations where interface deformations result from either bulk flows triggered by bulk forces~\cite{schroll_liquid_2007,wunenburger2011,chraibi_excitation_2013,chesneau2020dynamics,Tan2009} or thermocapillary flows originating from interfacial tension gradients triggered by temperature gradients~\cite{chraibi2012,riviere2016}.
Regarding non-scattering and non-absorbing liquids, the previously cited studies have led to conclude that the observed morphological diversity of fluid interface deformations originates basically from the mutual interplay between the shape of the interface deformed by the radiation pressure and the wave propagation. Indeed, a distorted interface acts as a dioptre that modifies the structure of the wave passing through it, hence the radiation pressure exerted on it. From now on, we refer to such a feedback mechanism as the propagation-deformation interplay (PDI), which is the subject of this work.

In the regime of `small' deformations (i.e., whose height is smaller than the beam diameter), the disturbances of the beam by the curved interface are noticeable only far downstream of the interface~\cite{casner_adaptative_2001}. Consequently, the shape of a static interface can be quantitatively described by solving the balance between Laplace pressure, buoyancy and the radiation pressure exerted by the incident, \red{\it non perturbed}, wave~\cite{Casner2001,Wunenburger2006,Issenmann2008}. By contrast, \red{the description of large interface deformations requires to account for the perturbation of the wave by the interface, i.e. PDI}. The particular case of axisymmetric liquid columns can be treated analytically owing to the translational invariance of the interface shape along its axis of revolution~\cite{Brasselet2008,Brasselet2008_2,Bertin2010}. The predicted equilibrium radius of the column, which has been calculated within an exact scalar description of the electromagnetic field in both liquids, is in agreement with experimental observations~\cite{Brasselet2008,Brasselet2008_2}. The case of stepped deformations is far more challenging from an analytical point of view. Still, the beam propagation through axisymmetric dioptres having the shape of the experimentally observed steady interfaces can be numerically computed. The radiation pressure distribution deduced from the computed wave field was found to satisfactorily balance the Laplace pressure and buoyancy assessed from the measured shape of the interface~\cite{Bertin2012}. Although this is an {\it a posteriori} verification of the validity of the PDI hypothesis, a self-consistent determination of the shapes of the irradiated interfaces resulting from the PDI is still missing. Furthermore, the mechanism underlying the instability leading to the formation of needle-like deformations, which is likely to involve PDI, is still unknown~\cite{Casner2003,Wunenburger2006}.

In this work we aim to determine the interface deformations by solving the PDI problem for any deformation amplitude. This is done by numerically mimicking an experiment. In other words, we simultaneously solve the hydrodynamic evolution of an irradiated two-phase fluid sample and the propagation of the acoustic or electromagnetic beam through the moving interface, taking a planar and horizontal interface at rest as the initial condition (time $t=0$) and considering a nonzero constant beam power for $t>0$. Such a \red{numerical} simulation of the radiation pressure-driven deformation morphogenesis of fluid interfaces allows us to address the following questions:

(i) How do the interface deformations depend on the various parameters characterizing the wave-matter interaction?

(ii) Do the numerical simulations predict the same interface deformation shapes as those experimentally observed?

(iii) Does the instability leading to the formation of needle-like deformations and the ensuing observed hysteresis originate from the PDI~\cite{Casner2003,Wunenburger2006}?

The paper is organized as follows. The key-ingredients of the model are presented in Section II. The dimensionless formulation of the model and its numerical handling are provided in Section III. Then, a dimensional analysis is performed in Section IV, which shows that the deformation height depends on five independent dimensionless quantities both in acoustics and electromagnetics. The numerical results and its comparison with existing and novel experimental observations are presented in Section V, which addresses the three main questions listed above. Finally, Section VI then summarizes the main results of this work.







\section{Model}\label{s:model}
In this section, we first present the wave propagation equations describing the acoustic and electromagnetic fields and the corresponding radiation stresses exerted on the interface. Then, we present the hydrodynamic equations describing the evolution of the fluid sample in the creeping-flow regime that suits well most of the experiments reported so far.


%
\begin{figure}[t!]
	\begin{center}
	\includegraphics[width=1\columnwidth]{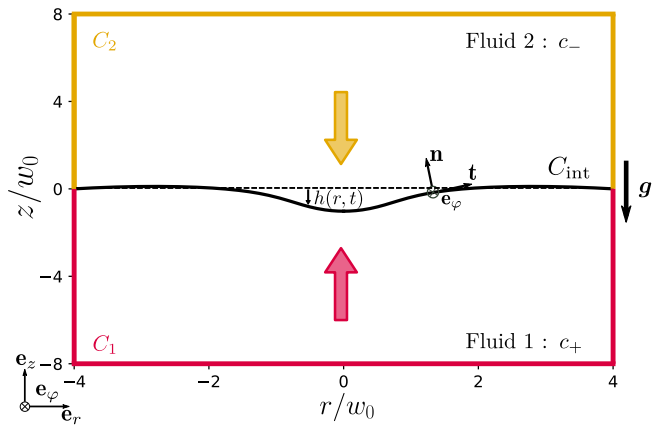}
	\caption{Sketch of the simulation domain with the definitions of the used notations. The convention $c_1=c_{+}$ and $c_2=c_{-}$ with $c_{-} < c_{+}$ is used, see Section V A for details. The bold vertical arrows refer to the two possible directions of propagation of the irradiating waves. \label{fig:sketch}}
	\end{center}
\end{figure}
%
%
\subsection{Propagation}\label{ss:propagation}
Considering both fluids $1$ and $2$ as homogeneous, isotropic and linear, we also assume inviscid and non heat-conducting media when dealing with acoustic waves whereas we assume non absorbing and non dispersive media for electromagnetic waves. Within the latter framework, the propagation of small-amplitude acoustic waves and of electromagnetic waves is described by d'Alembert's wave equation
\begin{equation}
\label{eq:helmholtz}
	\frac{\partial ^2 q_i}{\partial t^2} - c_i^2 \, \nabla^2 q_i = 0, \, i=(1,2),
\end{equation}  
where $q$ is the fluid pressure perturbation $p^\prime$ associated to acoustic waves, whereas it refers to the electric field $\mathbf{E}$ or the magnetic field $\mathbf{H}$ for electromagnetic waves, and $c$ is the wave celerity (also equal to the phase and group velocities). In addition the subscript $i$, refers to the fluid labeled $i$, $i=(1,2)$, see Fig.~\ref{fig:sketch}.
\subsubsection{Acoustic waves}
In the case of acoustic waves, $c_i=(\rho_i \chi_i)^{-1/2}$ where $\rho_i$ is the mass density and $\chi_i$ the isentropic compressibility. The complex harmonic velocity field $\mathbf{u}_i$ associated to the acoustic pressure perturbation $p^\prime_i$ satisfies the linearized Euler's equation $\rho_i \frac{\partial \mathbf{u}_i}{\partial t} = - \nabla p^\prime_i$. Along the fluid interface,
the pressure and the velocity fields satisfy the condition of stress continuity
\begin{equation} \label{eq:Contp}
    p^\prime_1=p^\prime_2\,,
\end{equation}
as well as the impermeability condition between the two fluids
\begin{equation}\label{eq:Contun}
    \mathbf{u}_1 \cdot \mathbf{n} = \mathbf{u}_2 \cdot \mathbf{n}\,,
\end{equation}
where $\mathbf{n}$ is the unit vector normal to the interface oriented from fluid $1$ to fluid $2$. Finally, since the acoustic fluid velocity is irrotational in absence of acoustic attenuation, $\mathbf{u}_i = - \nabla \phi_i$ and consequently $p^\prime_i =  \rho_i \frac{\partial \phi_i}{\partial t}$. Thus, the continuity of $p^\prime$ across the interface results in the continuity of $\rho \phi$, hence of $\nabla_{\|} (\rho \phi) = \rho \nabla_{\|} \phi$, where $\nabla_{\|}$ is the gradient in the tangent plane to the interface. Since $\nabla_{\|} \phi = - {\bf u}_{\|}$, where ${\bf u}_{\|} = {\bf u} - ({\bf u} \cdot {\bf n}) {\bf n}$ is the projection of ${\bf u}$ on the tangent plane to the interface, we conclude that
\begin{equation}\label{eq:Contut}
    \rho_1 \, {\bf u}_{\| 1} = \rho_2 \, {\bf u}_{\| 2}.
\end{equation}
\subsubsection{Electromagnetic waves}
In the case of electromagnetic waves, $c_i=(\varepsilon_i \mu_i)^{-1/2}$ where $\varepsilon_i$ is the dielectric permittivity and $\mu_{i}$ is the magnetic permeability (which is usually equal to its value in vacuum, $\mu_0$, in the optical frequency domain), both considered as real. The harmonic fields $\mathbf{E}_i$ and $\mathbf{H}_i$ satisfy Maxwell's equations $\nabla \cdot (\varepsilon_i \mathbf{E}_i) = \mathbf{0}$, $\nabla \times \mathbf{E}_i =  - \mu_i  \frac{\partial \mathbf{H}_i}{\partial t}$, $\nabla \cdot (\mu_i \mathbf{H}_i) = \mathbf{0}$ and $\nabla \times \mathbf{H}_i =  \varepsilon_i \frac{\partial \mathbf{E}_i}{\partial t}$. Thus, along the fluid interface, $\mathbf{E}_i$ and $\mathbf{H}_i$ satisfy
\begin{eqnarray}
\varepsilon_1 \mathbf{E}_1 \cdot \mathbf{n} &=& \varepsilon_2 \mathbf{E}_2 \cdot \mathbf{n}, \label{eq:ContEn}\\
\mathbf{n} \times \mathbf{E}_{1} &=& \mathbf{n} \times \mathbf{E}_{2}, \label{eq:ContEt}\\
\mu_1 \mathbf{H}_1 \cdot \mathbf{n} &=& \mu_2 \mathbf{H}_2 \cdot \mathbf{n}, \label{eq:ContHn}\\
\mathbf{n} \times \mathbf{H}_{1} &=& \mathbf{n} \times \mathbf{H}_{2}. \label{eq:ContHt}
\end{eqnarray}
The refractive index is $n_i=\sqrt{(\varepsilon_i \mu_i)/(\varepsilon_0 \mu_0)}=c_0/c_i$, where $\varepsilon_0$ is the dielectric permittivity of vacuum and $c_0=(\varepsilon_0 \mu_0)^{-1/2}$ is the wave celerity in vacuum. Note that in acoustics, a refractive index can be also defined as $n_i = 1/c_i$.


%
\subsection{Radiation stress}

In this section we define the acoustic and electromagnetic stresses exerted on the fluid interface, and we derive their expression assuming incident fields having axisymmetric incident intensity profiles. The beam propagation direction defines the $z$ axis, which is normal to the interface at rest. We thus introduce the cylindrical coordinates $(r,\varphi,z)$ associated with the direct orthonormal basis $(\mathbf{e}_r, \mathbf{e}_{\varphi}, \mathbf{e}_{z})$ and also the local direct orthonormal basis $(\mathbf{n}, \mathbf{t}, \mathbf{e}_{\varphi})$, see Fig.~\ref{fig:sketch}.

\subsubsection{Acoustic waves}
The time-averaged acoustic radiation stress $\mathbf{T}_{\rm R}^{\text{AC}}$ exerted on the fluid interface by the surrounding acoustic field is expressed as
\begin{equation}
  \mathbf{T}_{\rm R}^{\text{AC}} = \left( \mathbb{T}^{\text{AC}}_2 -  \mathbb{T}^{\text{AC}}_1 \right)\cdot \mathbf{n} ,
\end{equation}
where $\mathbb{T}^{\text{AC}}$ is the time-averaged acoustic radiation tensor defined in each fluid as~\cite{brillouin1938,borgnis1953}:
\begin{equation}\label{Brillouin_stress_tensor}
    	\mathbb{T}^{\text{AC}}_i = -\frac{1}{2} \chi_i \, \langle p_i^{\prime 2} \rangle \; \mathbb{I} + \frac{1}{2} \rho_i \langle \mathbf{u}_i^{2} \rangle \; \mathbb{I} - \rho_i \langle \mathbf{u}_i \otimes \mathbf{u}_i \rangle.
\end{equation}
$\mathbb{I}$ is the identify tensor, $\otimes$ is the dyadic product, and $\langle \cdot \rangle$ refers to the time average over one period $T = 2\pi/\omega$.

Assuming the incident acoustic pressure and velocity fields to have axisymmetric amplitude and phase spatial distribution, the acoustic velocity can be expressed as $\mathbf{u}_i = u_{\rmn i} \mathbf{n} + u_{\rmt i} \mathbf{t}$ and one finds for the normal component of the acoustic radiation stress 
\begin{equation}\label{eq:stressac}
\begin{split}
\mathbf{T}_{\rm R}^{\text{AC}} \cdot \mathbf{n} =& \left[ -\frac{1}{2} \chi_2 \, \langle p_2^{\prime 2} \rangle  
    + \frac{1}{2} \rho_2 \left( \langle u_{\rmt 2}^2 \rangle - \langle u_{\rmn 2}^2 \rangle \right) \right]\\
    &- \left[ -\frac{1}{2} \chi_1 \, \langle p_1^{\prime 2} \rangle  
    + \frac{1}{2} \rho_1 \left( \langle u_{\rmt 1}^2 \rangle - \langle u_{\rmn 1}^2 \rangle \right) \right]\,,
\end{split}
\end{equation}
where, according to Eqs.~\eqref{eq:Contun} and \eqref{eq:Contut}, the normal and tangential components of the velocity field on both sides of the interface satisfy 
\begin{eqnarray}
\rho_1 u_{\rmt 1} &=& \rho_2 u_{\rmt 2}\label{eq:Contut2},\\
u_{\rmn 1} &=& u_{\rmn 2}\label{eq:Contun2}.
\end{eqnarray}
Noticeably, the tangential component of the acoustic radiation stress $\mathbf{T}_{\rm R}^{\text{AC}} \cdot \mathbf{t}$ is zero, which justifies the improper name of radiation pressure given to the acoustic radiation stress although it is not isotropic~\cite{marston1980}. Since $\mathbf{T}_{\rm R}^{\text{AC}} \cdot \mathbf{n}$ is axisymmetric, we expect axisymmetric interface deformations, which the problem two-dimensional.

Considering an incident plane wave, the acoustic radiation pressure exerted on the horizontal interface at rest is
\begin{equation}
\label{eq:TR0AC}
\mathbf{T}_{{\rm R},0}^{\text{AC}} = \frac{2I}{c_{\text{from}}}\frac{Z_1^2 + Z_2^2 - 2 \frac{c_1}{c_2} Z_1 Z_2}{(Z_1 + Z_2)^2} \, \mathbf{n},
\end{equation}
where $I$ is the acoustic intensity of the incident plane wave, $c_{\text{from}}$ is the wave celerity of the medium from which the incident wave impinges on the interface and $Z_i=\rho_i c_i$ is the acoustic impedance~\cite{borgnis1953,Issenmann2008}. In particular, when $\rho_1=\rho_2$, Eq. \eqref{eq:TR0AC} simplifies to
\begin{equation}
\label{eq:TR0AC2}
\mathbf{T}_{{\rm R},0}^{\text{AC}} = \frac{2I}{c_{\text{from}}} \frac{c_2-c_1}{c_1 + c_2} \, \mathbf{n}.
\end{equation}

%
\subsubsection{Electromagnetic waves}
The time-averaged electromagnetic radiation stress $\mathbf{T}_{\rm R}^{\text{EM}}$ exerted on the fluid interface by the surrounding electromagnetic field is expressed as
\begin{equation}
    \mathbf{T}_{\rm R}^{\text{EM}} =  \left( \mathbb{T}^{\text{EM}}_2  - \mathbb{T}^{\text{EM}}_1 \right) \cdot \mathbf{n}\,,
\end{equation}
where $\mathbb{T}^{\text{EM}}$ is the time-averaged electromagnetic radiation tensor defined in each fluid as~\cite{Brevik1979}:
\begin{equation}	\label{Maxwell_stress_tensor}
\mathbb{T}^{\text{EM}}_i = -\frac{1}{2} \varepsilon_i \langle \mathbf{E}_i^{2} \rangle \;  \mathbb{I} - \frac{1}{2} \mu_{i} \langle \mathbf{H}_i^{2} \rangle \; \mathbb{I} + \varepsilon_i \langle \mathbf{E}_i \otimes \mathbf{E}_i \rangle + \mu_{i} \langle \mathbf{H}_i \otimes \mathbf{H}_i \rangle.
\end{equation}
The electrostriction term is purposely discarded since it does not contribute to the stress balance at the interface~\cite{Chraibi2008}.

Noteworthy, in contrast to the case of acoustic waves, it is not enough to assume axisymmetric phase spatial distribution for both the electric and  magnetic fields in order to get an axisymmetric radiation stress leading to a two-dimensional problem. Axisymmetric electromagnetic radiation stress can be actually obtained by considering either axisymmetric transverse electric (TE) or axisymmetric transverse magnetic (TM) polarized electromagnetic beams.

On the one hand, for a TE field defined as $\mathbf{E}_i = E_i \mathbf{e}_{\varphi}$ and $\mathbf{H}_i = H_{\rmn i} \mathbf{n} + H_{\rmt i} \mathbf{t}$, the normal component of the electromagnetic radiation stress is
\begin{equation}\label{eq:stressEMTE}
\begin{split}
\mathbf{T}_{\rm R}^{\text{EM,TE}} \cdot \mathbf{n} =& \left[ -\frac{1}{2} \varepsilon_2 \, \langle E_2^2 \rangle  
    + \frac{1}{2} \mu_2 \left( \langle H_{\rmn 2}^2  \rangle - \langle H_{\rmt 2}^2 \rangle \right) \right]\\
    & - \left[ -\frac{1}{2} \varepsilon_1 \, \langle E_1^2 \rangle  
    + \frac{1}{2} \mu_1 \left( \langle H_{\rmn 1}^2 \rangle - \langle H_{\rmt 1}^2 \rangle \right) \right],
\end{split}
\end{equation}
where, according to Eqs.~\eqref{eq:ContEn}--\eqref{eq:ContHt}, the magnitude of the electric field and the normal and tangential components of the magnetic field on both sides of the interface satisfy 
\begin{eqnarray}
E_1 &=& E_2 \label{eq:ContEn2},\\
\mu_1 H_{\rmn 1} &=& \mu_2 H_{\rmn 2}\label{eq:ContHn2},\\
H_{\rmt 1} &=& H_{\rmt 2}.\label{eq:ContHt2}
\end{eqnarray}
On the other hand, for a TM field defined as $\mathbf{E}_i = E_{\rmn i} \mathbf{n} + H_{\rmt i} \mathbf{t}$ and $\mathbf{H}_i = H_i \mathbf{e}_{\varphi}$, one gets 
\begin{equation}\label{eq:stressEMTM}
\begin{split}
\mathbf{T}_{\rm R}^{\text{EM,TM}} \cdot \mathbf{n} =& \left[ -\frac{1}{2} \mu_2 \, \langle H_2^2 \rangle  
    + \frac{1}{2} \varepsilon_2 \left( \langle E_{\rmn 2}^2  \rangle - \langle E_{\rmt 2}^2 \rangle \right) \right]\\
    & - \left[ -\frac{1}{2} \mu_1 \, \langle H_1^2 \rangle  
    + \frac{1}{2} \varepsilon_1 \left( \langle E_{\rmn 1}^2 \rangle - \langle E_{\rmt 1}^2 \rangle \right) \right],
\end{split}
\end{equation}
where, according to Eqs.~\eqref{eq:ContEn}--\eqref{eq:ContHt}, the normal and tangential components of the electric field and the magnitude of the magnetic field and on both sides of the interface satisfy 
\begin{eqnarray}
\varepsilon_1 E_{\rmn 1} &=& \varepsilon_2 E_{\rmn 2}\label{eq:ContEn3},\\
E_{\rmt 1} &=& E_{\rmt 2},\label{eq:ContEt3}\\
H_1 &=& H_2. \label{eq:ContHn3}
\end{eqnarray}
Moreover, the tangential component of the electromagnetic radiation stress $\mathbf{T}_R^{\text{EM}} \cdot \mathbf{t}$ is zero whatever the polarization state.

Considering an incident plane wave, the electromagnetic radiation pressure exerted on the horizontal interface at rest is independent of the polarization state and equals
\begin{equation}
\label{eq:TR0EM}
    \mathbf{T}_{{\rm R},0}^{\text{EM}} = \frac{2I}{c_{\text{from}}} \frac{Z_1^2 + Z_2^2 - 2 \frac{c_1}{c_2} Z_1 Z_2}{(Z_1 + Z_2)^2}\, \mathbf{n},
\end{equation}
where $I$ is the electromagnetic intensity of the incident plane wave, $c_{\text{from}}$ is the wave celerity of the medium from which the incident wave impinges on the interface and $Z_i=(\mu_i / \varepsilon_i)^{1/2} =(\varepsilon_i c_i)^{-1} = \mu_ic_i$ is the electromagnetic impedance.
In particular, in the optical frequency domain for which $\mu_1=\mu_2$, Eq. \eqref{eq:TR0EM} simplifies to
\begin{equation}
\label{eq:TR0EM2}
\mathbf{T}_{{\rm R},0}^{\text{EM}} = \frac{2I }{c_{\text{from}}} \frac{c_2-c_1}{c_1 + c_2} \,  \mathbf{n}\,.
\end{equation}
where $I$ is the electromagnetic intensity of the incident plane wave.
\subsection{Synthesis}
Comparisons between Eqs.~\eqref{eq:Contp} and \eqref{eq:ContEn2}, between  Eqs.~\eqref{eq:Contut2} and \eqref{eq:ContHn2} between Eqs.~\eqref{eq:Contun2} and \eqref{eq:ContHt2} and between Eqs.~\eqref{eq:stressac} and \eqref{eq:stressEMTE} show that the electromagnetic TE problem and the acoustic problem are formally identical with the following correspondences: $p \leftrightarrow E$, $\mathbf{u} \leftrightarrow  \mathbf{e}_{\varphi} \times \mathbf{H}$, $\rho \leftrightarrow  \mu$ and $\chi \leftrightarrow  \varepsilon$. Similarly, comparisons between Eqs.~\eqref{eq:Contp} and \eqref{eq:ContHn3}, between  Eqs.~\eqref{eq:Contut2} and \eqref{eq:ContEn3}, between Eqs.~\eqref{eq:Contun2} and \eqref{eq:ContEt3} and between Eqs.~\eqref{eq:stressac} and \eqref{eq:stressEMTM} show that the electromagnetic TM problem and the acoustic problem are formally identical with the following correspondences: $p \leftrightarrow H$, $\mathbf{u} \leftrightarrow  \mathbf{e}_{\varphi} \times \mathbf{E}$, $\rho \leftrightarrow  \varepsilon$ and $\chi \leftrightarrow  \mu$. 

These correspondences between the mechanical effects of acoustic waves and TE or TM electromagnetic waves offer a common framework that allow treating the wave-matter interaction in a universal manner.

%
\subsection{Flow}
As a result of the radiation stresses exerted on the interface, the interface can be deformed and the two-phase fluid sample can flow. The velocity and pressure fields associated to this flow driven by radiation pressure are noted $\mathbf{U}_i$ and $P_i$, respectively. 

Since the observed interface velocity is always largely subsonic, we assume both velocity fields $\mathbf{U}_i$ as incompressible, hence satisfying
\begin{equation}\label{eq:incomp}
   \nabla \cdot \mathbf{U}_i = 0\,.
\end{equation}
and both fluid densities $\rho_i$ remain homogeneous and constant.

Moreover, the observed interface velocity is sufficiently small to ensure that the associated Reynolds number is small compared to unity and we model the flows of both fluids as creeping flows. Defining the piezometric pressure $\hat{P}_i$ as $\hat{P}_i = P_i + \rho_i g z$, where $\mathbf{g} = - g\mathbf{e}_z$ is the gravity acceleration, Stokes equation is therefore satisfied in both fluids~\cite{Chraibi2008}, namely,
\begin{equation}\label{eq:stokes}
    -\nabla \hat{P}_i + \eta_i \,  \nabla^{2}\mathbf{U}_i = 0\,,
\end{equation}
where $\eta_i$ is dynamic viscosity of fluid $i$. In addition, at the interface between the two immiscible fluids, we assume the continuity of the flow velocity
\begin{equation}\label{eq:condcin}
    \mathbf{U}_1 = \mathbf{U}_2\,.
\end{equation}

The hydrodynamic stress $\mathbf{T}_{\rm H0}$ exerted by the flows on the interface is
\begin{equation}
    \mathbf{T}_{\rm H0} = \left( \mathbb{T}_{2}^{\text{H0}} - \mathbb{T}_{1}^{\text{H0}} \right)  \cdot \mathbf{n}.
\end{equation}
where $\mathbb{T}^{\text{H0}}_i$ is the hydrodynamic stress tensor
\begin{equation}
   \mathbb{T}^{\text{H0}}_i = - P_i \; \mathbb{I} +  \frac{\eta_i}{2} \left[ \nabla \mathbf{U}_i + (\nabla \mathbf{U}_i)^{\mathsf{T}}\, \right]\,,
\end{equation}
$(\cdot)^{\mathsf{T}}$ refering to the transpose. 

Along the moving interface, the sum of the hydrodynamic and radiation pressures stresses is balanced by the Laplace pressure:
\begin{equation}
\label{eq:balancepresquefinal}
	 \mathbf{T}_{\rm H0} + \mathbf{T}_{\rm R} + \sigma \kappa \, \mathbf{n} = 0.\end{equation}
where $\sigma$ is the interfacial tension, $h(r,t)$ is the height of the axisymmetric interface deformation at radius $r$ and time $t$ and
%
%
%
\begin{equation}
    \kappa = \frac{\partial h}{\partial s} \left( \frac{1}{r}  + \frac{\partial^2 h}{\partial s^2}\right) - \frac{\partial^2 r}{\partial s^2}
\end{equation}
is the associated curvature, $s$ being the curvilinear abscissa defined along the deformation $h(r,t)$, see Fig.~\ref{fig:sketch}.

In order to separate the effect of gravity from the effect of flows on the interface shape, it is useful to define the pseudo-stress tensor as
\begin{equation}
   \mathbb{T}^{\text{H}}_i = - \hat{P}_i \; \mathbb{I} +  \frac{\eta_i}{2} \left[ \nabla \mathbf{U}_i + (\nabla \mathbf{U}_i)^{\mathsf{T}}\, \right]\,
\end{equation}
and the hydrodynamic pseudo-stress as
\begin{equation}
    \mathbf{T}_{\rm H} = \left( \mathbb{T}_{2}^{\text{H}} - \mathbb{T}_{1}^{\text{H}} \right)  \cdot \mathbf{n}.
\end{equation}
Eq.~\eqref{eq:balancepresquefinal} can be rewritten as:
\begin{equation}
\label{eq:balancefinal}
	 \mathbf{T}_{\rm H} + \mathbf{T}_{\rm R} + \sigma \kappa \, \mathbf{n} + (\rho_1 - \rho_2) g h \, \mathbf{n} = 0.
\end{equation}
where $\mathbf{T}_{\rm H}$ solely encompasses all the effect of flows on the interface shape. 


\section{Numerical resolution}
%
\subsection{Dimensionless formulation}
The axisymmetry of the problem results in a two-dimensional formulation. As depicted in  Fig.~\ref{fig:sketch}, the computation domain is included in the meridional plane $(r,z)$ and is defined as $0 \leq r \leq r_{\rm max}$, $0 \leq z \leq z_{\rm max}$ (we chose $r_{\rm max} = 4 \omega_0$ and $z_{\rm max} = 8 \omega_0$). Accordingly, fluids $1$ and $2$ are associated with identical areas equal to $r_{\rm max}z_{\rm max}$ and whose closed-path contour is made of a fixed contour $C_i$, $i=(1,2)$, and a mobile contour $C_{\text{int}}$ which defines the fluid interface, as sketched in Fig.~\ref{fig:sketch}. The interface coincides with the plane $z=0$ at rest and its contact line is pinned at $(r=\pm r_{\rm max}, z=0)$ with variable contact angle.

The goal of the numerical simulation is to evaluate $\mathbf{U}$ and $P$ at any point $\mathbf{x}(r,z) = r\,\mathbf{e}_r + z\,\mathbf{e}_z$ of the computation domain and at any time $t \geq 0$ in order to compute the stresses exerted on the interface at its location, then its motion, which is computed using a Lagrangian approach according to
\begin{equation}\label{eq:lagrange}
    \frac{\text{d} {\bf x}}{\text{d} t} = \mathbf{U}(\mathbf{x}) \, , \, {\bf x} \in C_{\text{int}},
\end{equation}
starting from rest at $t=0$.

The incident acoustic or electromagnetic beam is chosen to be focused at the fluid interface at rest, as is usually done in experiments. More precisely, based on an experimental argument, we choose Gaussian incident beams whose intensity profiles at $z=0$ in absence of interface is
\begin{equation}
\label{eq:GaussianIntensity}
    I(r)= \frac{2{\cal P}}{\pi w_0^2} \exp (-2r^2/w_0^2)\,,
\end{equation}
where ${\cal P}$ is the total beam power. Indeed, in the electromagnetic case, a continuous-wave laser sources in the $\text{TEM}_{00}$ mode is usually used, see for instance~\cite{Casner2001,Wunenburger2006}. In the case of acoustics, spherically focused single-element transducers are usually used~\cite{Issenmann2008,Bertin2012}. Still, it has been shown in~\cite{issenmann2006} that Eq.~\eqref{eq:GaussianIntensity} accurately fits the acoustic intensity distribution in the focal plane of these transducers. To conclude, Eq.~\eqref{eq:GaussianIntensity} is adapted to the description of electromagnetic and acoustic beams at their focus.

Using $w_0$ as the characteristic spatial scale we can define the characteristic fluid velocity in the creeping flow regime as $U_{0} = \sigma/\eta_2$ (note: choosing $\eta_1$ instead of $\eta_2$ would be equally relevant), the characteristic interface evolution timescale as $\tau = w_{0}/U_{0}$ and the characteristic pseudo-pressure variation in each fluid as $\hat{P}_{i0} = \eta_{i} U_{0}/w_{0}$. These characteristic parameters allow to formulate the problem in a dimensionless manner. namely, we introduce the dimensionless position vector $\tilde{\mathbf{x}} = \mathbf{x}/w_0$, time $\tilde{t}=t/\tau$, velocity $\tilde{\mathbf{U}}_i = \mathbf{U}_i/U_{0}$, pseudo-pressure $\tilde{P}_i=\hat{P}_i/\hat{P}_{i0}$ and deformation height $\tilde{h}=h/w_0$. Accordingly, Eqs.~\eqref{eq:incomp}--\eqref{eq:condcin} and \eqref{eq:balancefinal} are rewritten in a dimensionless form as
\begin{gather}
	\tilde{\nabla} \cdot \tilde{\mathbf{U}}_{i} = 0\,,\label{eq:incompadim}\\
	-\tilde{\nabla} \tilde{P}_{i} + \tilde{\nabla}^{2}\tilde{\mathbf{U}}_{i} = 0\,, \label{eq:stokesadim}\\
	\tilde{\mathbf{U}}_{1} = \tilde{\mathbf{U}}_{2}  \, , \, {\bf x} \in C_{\text{int}}\,,\\
		( \tilde{\mathbb{T}}_{2}^{\text{H}}-\alpha\tilde{\mathbb{T}}_{1}^{\text{H}}) \cdot \mathbf{n} +  \tilde{\mathbf{T}}_{R} +
		 \tilde{\kappa} \mathbf{n} +   \text{Bo} \, \tilde{h} \mathbf{n} =0  \, , \, {\bf x} \in C_{\text{int}}\,,
\end{gather}
where $\alpha = \eta_{1}/\eta_{2}$, $\text{Bo} = (\rho_{1}-\rho_{2}) g w_{0}^{2} / \sigma$ is the Bond number, $\tilde{\mathbb{T}}_i = \mathbb{T}_i/\hat{P}_{i0}$ is the dimensionless hydrodynamic stress tensor and $\tilde{\mathbf{T}}_{R} = \mathbf{T}_{R} \, w_{0} / \sigma$ is the dimensionless radiation stress. Finally, the boundary condition along $C_1$ and $C_2$ is
\begin{equation}
    \tilde{\mathbf{U}}_{i} = \mathbf{0}   \, , \, {\bf x} \in C_i\,.
\end{equation}
\subsection{Boundary Integral formulation}
The solution of Eqs.~\eqref{eq:incompadim} and \eqref{eq:stokesadim} can be obtained by determining the solution of the Stokes equation corresponding to a Dirac-excitation~\cite{Ladyzhenskaya1987,Pozrikidis1992}. It consists in exerting a unit point-force on the fluid at a given point and determining the induced velocity field $\tilde{\mathbf{U}}$ and hydrodynamic stress tensor $\mathbb{\tilde{T}}^{\text{H}}$ everywhere else.
Taking advantage of axisymmetry of the problem, the solution can be eventually expressed as a function of one-dimensional integrals along the contours $C_1$, $C_2$ and $C_{\rm int}$, see Ref.~\cite{Chraibi2010} for details. One finds
\begin{equation}
	\begin{split}
	&\frac{1+\alpha}{2} \,  \tilde{\mathbf{U}}(\tilde{\bf x}) =\\ & - \int_{C_{\text{int}}} \! \! \! \! \!   \left[ \tilde{\kappa}(\tilde{\bf y}) + \tilde{T}_{R}(\tilde{\bf y}) + \text{Bo} \, \tilde{h}(\tilde{\bf y})\right] \mathbb{\tilde{U}}^{*}(\tilde{\bf x}-\tilde{\bf y}) \cdot \mathbf{n}(\tilde{\bf y}) \; \text{d}C(\tilde{\bf y})\\
	 & + (1-\alpha) \int_{C_{\rm int}} \! \! \! \! \! \left[\mathbb{\tilde{K}}^{*}({\bf \tilde{x}}-{\bf \tilde{y}})\cdot
	 \mathbf{\tilde{U}}(\tilde{\bf y})\right] \cdot \mathbf{n}(\tilde{\bf y}) \,  \text{d}C(\tilde{\bf y})\\
	 & - \alpha \int_{C_{1}}  \mathbb{\tilde{U}}^{*}({\bf \tilde{x}}-{\bf \tilde{y}}) \cdot
	 \left[\mathbb{\tilde{T}}_{1}^{\text{H}}(\mathbf{\tilde{y}}) \cdot \mathbf{n}(\tilde{\bf y})\right] \, \text{d}C(\tilde{\bf y}) \\
	 & +\int_{C_{2}} \mathbb{\tilde{U}}^{*} ({\bf \tilde{x}}-{\bf \tilde{y}}) \cdot \left[ \mathbb{\tilde{T}}_{2}^{\text{H}}(\tilde{\bf y}) \cdot \mathbf{n}(\tilde{\bf y})\right] \, \text{d}C(\tilde{\bf y})
	 \, , \, \tilde{\bf x} \in C_{\rm int}\,,
	\end{split}
	\label{integral_form}
\end{equation}
where $dC(\tilde{\bf y})$ refers to the elementary contour piece at point $\tilde{\bf y}$, and 
\begin{align}
	\mathbb{\tilde{U}}^{*}(\tilde{\bf d}) =& \frac{1}{8\pi}\Bigg(\frac{1}{\tilde{d}} \mathbb{I} + \frac{ \tilde{\bf d} \otimes \tilde{\bf d} }{\tilde{d}^{3}}\Bigg)\,,\\
	\mathbb{\tilde{K}}^{*}(\tilde{\bf d}) =& -\frac{3}{4\pi} \Bigg(\frac{\tilde{\bf d} \otimes \tilde{\bf d} \otimes \tilde{\bf d} }{\tilde{d}^{5}}\Bigg)\,.
\end{align}
are the velocity and stress associated to the Green kernel, respectively, with $\tilde{\bf d} = \tilde{\bf x} - \tilde{\bf y}.$

The numerical resolution is based on the Boundary Element Method. This method has been chosen because it allows for a precise monitoring of the interface position, which is crucial here since both capillary stresses and radiation pressure depend on the interface shape (i.e., slope and curvature). 
\subsection{Numerical procedure}
\subsubsection{Propagation}
The acoustic or electromagnetic beam is considered as harmonic with angular frequency $\omega$. As a consequence, the propagating field  $q_i =p^\prime_i$, $\mathbf{E}_i$ or $\mathbf{H}_i$ can be described as the real part of the complex field $Q_i \exp(-j \omega t)$ which satisfies Helmholtz's equation
\begin{equation}
\label{eq:helmholtz2}
	\nabla^{2} Q_i + k_i^{2} Q_i = 0,
\end{equation} 
where $k_i= \omega/c_i$ is the wavenumber in fluid $i$. Thus, the open source numerical code developed for computing the acoustic propagation in the harmonic regime~\cite{programmesKirkup} can be used indistinctly for simulating optical or acoustic fields. 
Eq.~\eqref{eq:helmholtz2} is solved using opensource \textsc{fortran} routines~\cite{programmesKirkup} described in Ref.~\cite{Kirkup2007} and adapted to our specific problem.

\subsubsection{Flow}
The influence of the viscosity ratio $\eta_{1}/\eta_{2}$ on the interface dynamics has been studied in~\cite{Chraibi2010}. The conclusion of the latter work is that the viscosity ratio has no qualitative influence on it, hence we chose $\eta_{1} = \eta_{2} = \eta$ for our numerical investigations.

First, the acoustic or electromagnetic field along the initially flat interface is computed. Then, the radiation pressure exerted along the interface is computed and injected into the first term of the right-hand side of Eq.~\eqref{integral_form}. 
This gives a system of $N$ integral equations, $N$ being the number of points used to discretize the interface.
Its numerical solution  
allows the determination of the velocity field along the interface, whose position is then advected
through a simple Euler forward scheme $\mathbf{x}(t+\Delta t) = \mathbf{x}(t) + \mathbf{U}(t) \Delta t$. This procedure is then repeated accounting for  until a stationary state is reached.
%
\section{Preliminary analysis}
%
\subsection{Universal dimensionless characteristic quantities}
In this section, we introduce two dimensionless quantities enabling, on the one hand, to evaluate the strength of the mechanical perturbation applied on the interface and on the other hand, to appreciate the influence of the deformed interface acting as a dioptre on the wave propagation. These two parameters are the dimensionless characteristic radiation pressure $\Pi_{0}$ and the characteristic waveguiding parameter $V$, respectively. Both are universal in the sense that they are independent of the acoustic or electromagnetic nature of the irradiating wave.

Constructing $\Pi_{0}$ starts by noting that a dimensionless radiation pressure $\Pi$ can be defined at each point of the interface as the ratio between the radiation pressure and the characteristic value of the Laplace overpressure $\sigma/w_0$,
\begin{equation}\label{eq:PiR}
\Pi = \frac{w_0}{\sigma}\, \mathbf{T}_R \cdot {\bf n}  \,,
\end{equation}
where $\mathbf{T}_R=\mathbf{T}_R^{\text{AC}}$ or $\mathbf{T}_R^{\text{EM}}$. A characteristic value of the dimensionless radiation pressure is thus obtained by considering a flat horizontal interface and evaluating it at the beam focus, see Eqs.~\eqref{eq:TR0AC} and \eqref{eq:TR0EM},
\begin{equation}\label{eq:PiR0}
    \Pi_{0} = \frac{2P}{\pi c_{\text{from}}\sigma w_0} \frac{Z_1^2 + Z_2^2 - 2 \frac{c_1}{c_2} Z_1 Z_2}{(Z_1 + Z_2)^2}.
\end{equation}
As shown in~\cite{Casner2001,Wunenburger2006,issenmann2006,Issenmann2008,Bertin2012}, in the $\text{Bo} \ll 1$ limit, which applies to the present work as discussed in Section V, the height $h_0$ of the steady humps formed at an interface irradiated by a Gaussian beam satisfies $h_0/w_0 \sim \Pi_{0}$.

The parameter $V$ is that arising from standard waveguide theory. Its introduction in the context of the present work is rooted in the fact that the existence and the diameter of liquid bridges~\cite{Brasselet2008,Brasselet2008_2,Bertin2010} and needle-like interface deformations~\cite{wunenburger2011} sustained either by acoustic or electromagnetic waves can be explained by the waveguiding properties of the wave-induced deformation that behaves as a step-index liquid-core liquid-cladding cylindrical waveguide. Considering the textbook case of step-index optical fibre with radius $R$, inner refractive index $n_{\rm in}$ and outer index $n_{\rm out} < n_{\rm in}$, the number of guided propagating modes is determined by the value of the dimensionless parameter $V = k_0 R \sqrt{n_{\rm out}^2 - n_{\rm in}^2}$, where $k_0$ is the wavenumber in vacuum of the harmonic field injected into the waveguide~\cite{okamoto2021}. By analogy, here we define a universal characteristic waveguiding parameter $V$ associated to an interface deformation as
\begin{equation}
\label{eq:V}
    V = \omega w_0 \sqrt{ \left| \left( \frac{1}{c_{\text{from}}} \right)^2 - \left( \frac{1}{c_{\text{to}}} \right)^2 \right|}\,,
\end{equation}
where $c_{\text{to}}$ is the wave celerity of the medium to which the incident wave goes after passing through the interface. Introducing the wavenumber of the incident beam, $k_{\text{from}}=\omega/c_{\text{from}}$, Eq.~\eqref{eq:V} can be reformulated as
\begin{equation}
\label{eq:V2}
V = k_{\text{from}} w_0 \sqrt{\left| 1 - \left( \frac{c_{\text{from}}}{c_{\text{to}}} \right)^2 \right|}\,.
\end{equation}
As shown in~\cite{Bertin2012}, $V$ is a prime choice parameter for describing and understanding the shape of observed interface deformations whatever is the nature of the wave.

Noteworthy, $\Pi_{0}$ and $V$ are the two key universal ingredients when addressing the PDI problem. Indeed, $\Pi_{0}$ informs to what extent the incident wave can deform the interface whereas $V$ informs to what extent the deformed interface can distort the incident wave.
%
\subsection{Dimensional analysis}
In this section we aim at identifying a minimal set of dimensionless parameters determining the axisymmetric shape of the deformed interface at steady state $h_\infty(r) = h(r,t \gg \tau)$.
\subsubsection{General remarks}
At first we notice that, for electromagnetic waves, the hypothesis of axisymmetric deformations requires to discard the role of the polarization state, which is valid in the limit of small refractive index contrast, as suggested by the expressions of the Fresnel coefficients of reflexion and transmission at the interface between two homogeneous media. Since all the reported electromagnetic experiments involve weak refractive index contrasts (typically of the order of a few percent), the role of the polarization state is discarded from now on. Moreover, the dimensions of the fluid sample $r_{\rm max}$ and $z_{\rm max}$ are assumed to be large enough that $h_{\infty}$ does not depend on them.

At steady state there is no more flow: ${\bf U}_i=\bf{0}$ and $\hat{P}_i=0$. Consequently $\mathbf{T}_{\rm H} = \mathbf{0}$. As shown by Eq.~\eqref{eq:balancefinal}, the deformation thus results from the combined effects of interfacial tension $\sigma$, buoyancy $g(\rho_1-\rho_2)$ and radiation pressure ${\bf T}_R$. As shown by Eq.~\eqref{Brillouin_stress_tensor} 
and ~\eqref{Maxwell_stress_tensor}, 
radiation pressure involves the propagating fields and physical properties of both fluids. The propagating fields are defined by the beam power ${\cal P}$, its waist $w_0$, its wavenumber $k_{\text{from}} = \omega/c_{\text{from}}$ and its direction of propagation, that is accounted by introducing a boolean variable $\xi$ whose value depends on whether $c_{\text{from}} > c_{\text{to}}$ or $c_{\text{from}} < c_{\text{to}}$. The fluids physical properties determining the radiation pressure are $c_1$, $c_2$ plus (i) $\mu_1$ and $\mu_2$ for electromagnetic waves or (ii) $\rho_1$ and $\rho_2$ for acoustic waves. In what follows we consider separately the acoustic and electromagnetic cases.


\subsubsection{Acoustic waves}

From the previous analysis it follows that for acoustic waves, $h_\infty$ is function of $11$ independent quantities:
\begin{equation}
\label{eq:ADac1}
    h_\infty = f(r, g(\rho_1-\rho_2), \sigma, {\cal P}, w_0, k_{\text{from}}, \xi, \rho_1, \rho_2, c_1, c_2).
\end{equation}
Since the quantities appearing in Eq.~\eqref{eq:ADac1} can be expressed using the dimensions of mass, length and time, according to Buckingham's theorem~\cite{buckingham1914}, Eq.~\eqref{eq:ADac1} can be rewritten as a relationship between 9 independent dimensionless quantities. We choose
\begin{equation}
\label{eq:ADac2}
    \frac{h_\infty}{w_0} = F \left( \frac{r}{w_0}, \text{Bo}, \Pi_{0}, V, \xi, \frac{Z_1}{Z_2}, \frac{c_1}{c_2},  A(\bar\rho) \right),
\end{equation}
where $\bar\rho = (\rho_1+\rho_2)/2$ and $A(\bar\rho)$ is a dimensionless quantity involving $\bar\rho$. Noticing that $\bar\rho$ accounts for fluid inertia, which is involved only in unsteady phenomena, $A(\bar\rho)$ is discarded recalling that we deal here with static steady states.
%
%
Finally, an important additional simplification occurs when considering $\rho_1=\rho_2$, which leads to $\text{Bo}=0$ and $Z_1/Z_2 = c_1/c_2$, hence to
\begin{equation}
\label{eq:ADac4}
    \frac{h_\infty}{w_0} = F \left( \frac{r}{w_0}, \frac{c_1}{c_2}, \Pi_{0}, V, \xi \right).
\end{equation}

\subsubsection{Electromagnetic waves}

In the case of electromagnetic waves, $h_\infty$ is function of $11$ independent quantities:
\begin{equation}
\label{eq:ADem1}
    h_\infty = f(r, g(\rho_1-\rho_2), \sigma, {\cal P}, w_0, k_{\text{from}}, \xi, \mu_1, \mu_2, c_1, c_2).
\end{equation}
The quantities appearing in Eq.~\eqref{eq:ADem1} can be expressed using the dimensions of mass, length, time, and electric current.
%
%
According to Buckingham's theorem, Eq.~\eqref{eq:ADem1} can be rewritten as a relationship between $8$ independent dimensionless quantities, which we choose as
\begin{equation}
\label{eq:ADem3}
    \frac{h_\infty}{w_0} = F \left( \frac{r}{w_0}, \text{Bo}, \Pi_{0}, V,  \xi, \frac{Z_1}{Z_2}, \frac{c_1}{c_2} \right).
\end{equation}
Taking into account that $\mu_1=\mu_2=\mu_0$  at optical frequencies and that $Z_i = \mu_i c_i$, and assuming $\rho_1=\rho_2$, one gets
\begin{equation}
\label{eq:ADem4}
    \frac{h_\infty}{w_0} = F \left( \frac{r}{w_0}, \frac{c_1}{c_2}, \Pi_{0}, V, \xi \right).
\end{equation}
%


\subsubsection{Synthesis}

This analysis shows that whenever the role of the polarization state of electromagnetic waves can be discarded, the steady dimensionless shape of the interface can be described in a universal manner, indistinctly for acoustic and electromagnetic waves. Moreover, in the $\rho_1 = \rho_2$ approximation, this universal description entails four dimensionless parameters, namely $\frac{c_1}{c_2}$, $\Pi_{0}$, $V$ and $\xi$. 

In the unsteady regime, $h$ also depends on time $t$, fluid viscosities $\eta_1$ and $\eta_2$, as well as $\rho_1$ and $\rho_2$. According to~\cite{Chraibi2010}, in the creeping flow regime, neither fluid inertia, nor the viscosity ratio have a significant influence on the interfacial dynamics, whose characteristic timescale is $\tau = w_0 \eta_2/\sigma$ as already introduced in section III.A.

\section{Results}
\subsection{Disclaimer}
With the aim at demonstrating that the PDI based on the sole action of radiation pressure presented in Section~\ref{s:model} reproduces the main features of experimental observations, we do not report here on the systematic exploration of the four-dimensional parameter space. Instead, we set $\Pi_{0}$ to a rather large value (typically, $\Pi_{0} \simeq 5-10$) in order to put the system in regimes where the consequences of PDI are strikingly apparent. Moreover, we focus on the influence of the celerity contrast and of the beam direction of propagation. For this aim, we choose two values that are representative of low and large values of the ratio $c_-/c_+$, namely, $c_-/c_+=0.5$, which can be achieved in acoustic experiments, and $c_-/c_+=0.95$, which is typical of optical experiments.

Moreover, the restrictions of our numerical study call for a few comments regarding a possible comparison with experimental results. Indeed, in our numerical study gravity effects are discarded owing to the assumption $\rho_1 = \rho_2$ while in practice $\rho_1 \neq \rho_2$. In the optical domain previous works showed that gravity does not have a qualitative influence on the observed deformations~\cite{Casner2001,Wunenburger2006} whereas in acoustics $\rho_1 \neq \rho_2$ results in a possible effect of gravity and also in an additional dimensionless parameter, the ratio $Z_1/Z_2$, as discussed in Section IV B. Therefore one can anticipate discrepancies between simulations and experiments in acoustics. 

\subsection{Morphogenesis of interface deformations: the role of propagation-deformation interplay}
In order to appreciate the role of PDI, we compare the deformation evolution of an initially flat interface experiencing constant irradiation from time $t=0$ with and without PDI. The case ``with PDI'' refers to the treatment presented so far whereas the case ``without PDI'' is carried out by computing the radiation pressure as that exerted by the incident Gaussian field whatever the deformation of the interface.

In order to appreciate the role of the direction of propagation, the comparison is conducted in two configurations, namely $c_{\text{from}} > c_{\text{to}}$ and $c_{\text{from}} < c_{\text{to}}$. In the following, we adopt the following convention: the bottom fluid indexed by the subscript $1$ is taken as that having the largest celerity ($c_1 = c_+$) whereas the top fluid indexed by the subscript $2$ has the smallest celerity ($c_2= c_-$ with $c_- < c_+$). Thus, the configuration $c_{\text{from}} > c_{\text{to}}$ corresponds to a beam propagating upward from fluid $1$ to fluid $2$. Conversely, the configuration $c_{\text{from}} < c_{\text{to}}$ corresponds to a beam propagating downward from fluid $2$ to fluid $1$.

\subsubsection{Large celerity contrast with $c_{\rm{from}} > c_{\rm{to}}$}
%
%

\begin{figure*}
\includegraphics[width=0.73\textwidth]{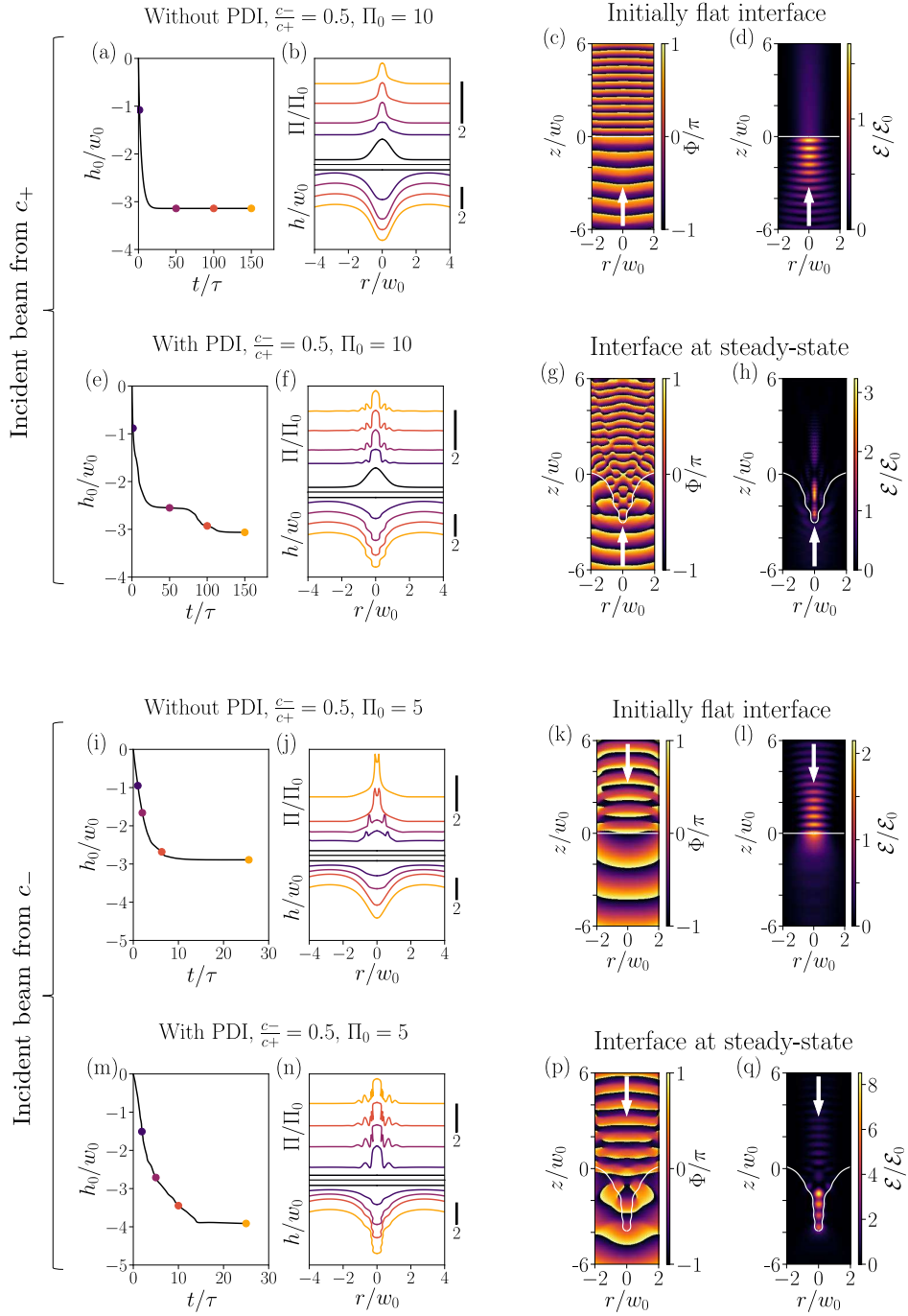}
\caption{\label{fig:dynamique1}
\scriptsize{Computed morphogenesis of large interface deformations for celerity contrast $c_{-}/c_{+} = 0.5$, waveguiding parameter $V = 5.4$ with Propagation-Deformation Interplay (PDI) (a,b, i, j) and without PDI (e)--(h), (m--q). (a)--(h) $c_{\text{from}} > c_{\text{to}}$, $\Pi_0 = 10$. (i)--(q) $c_{\text{from}} < c_{\text{to}}$, $\Pi_0 = 5$. The deformation profiles shown in panel (b) [respectively (f), (j) and (n)] correspond to the instants indicated by the symbols with corresponding color in panel (a) [respectively (e), (i) and (m)]. The curves in panels (b), (f), (j) and (n) are shifted vertically one with respect to the other for clarity purpose. In panels (c), (g), (k) and (p), $\Phi$ is the phase of the wave field. In panels (d), (h), (l) and (q), $\mathcal{E}$ is the wave energy density and $\mathcal{E}_0$ is the energy density maximum of the incident beam in absence of interface.}}
\end{figure*}

The evolution of the interface shape and of the radiation pressure distribution along the interface for $c_-/c_+=0.5$ and $c_{\text{from}} > c_{\text{to}}$, at $\Pi_{0} = 10$, is shown without PDI in Figs.~\ref{fig:dynamique1}(a, b) and with PDI in Figs.~\ref{fig:dynamique1}(e, f). An insight into the wave field if provided in the right part of the figure where the spatial distribution of the phase $\Phi$ and of the dimensionless energy density $\mathcal{E}/\mathcal{E}_{0}$ of the wave field, where $\mathcal{E}_{0}$ refers to the maximum of the energy density of the incident beam in absence of interface (i.e., considering that the fluid `from' fills all space). The phase and energy density at time $t=0$ (at which the interface is flat) are shown in Figs.~\ref{fig:dynamique1}(c) and \ref{fig:dynamique1}(d), respectively. Note the expected factor 2 between the wavelengths in the two fluids [Fig.~\ref{fig:dynamique1}(c)] and the axial energy modulation in fluid `from' [Fig.~\ref{fig:dynamique1}(d)] which results from the interference between the incident beam and the backward wave due to its partial reflection off the interface. 

In the absence of PDI, the magnitude of the dimensionless hump height $\tilde{h}_0=h_0/w_0$ smoothly increases with time and is associated with a bell shaped dimensionless hump profile $\tilde{h}(\tilde{r},\tilde{t})$ at all stages of the evolution, see Fig.~\ref{fig:dynamique1}(a), in agreement with previous studies~\cite{Chraibi2008,Chraibi2010}. Correspondingly, the dimensionless radiation pressure distribution along the deformed interface, $\Pi(\tilde{r},\tilde{t})$, is also smooth and bell-shaped close to the axis at all times.

By contrast, in presence of PDI, we observe a stepped evolution of the hump height and a transition from bell-shaped to step-shaped hump profiles, see Fig.~\ref{fig:dynamique1}(e). This transition is associated with the appearance of radial modulations of the radiation pressure exerted along the interface, whose physical origin lies in the appearance of guided modes within the increasingly slender deformation due to the PDI~\cite{Bertin2012}. 

\begin{figure*}
\includegraphics[width=0.75\textwidth]{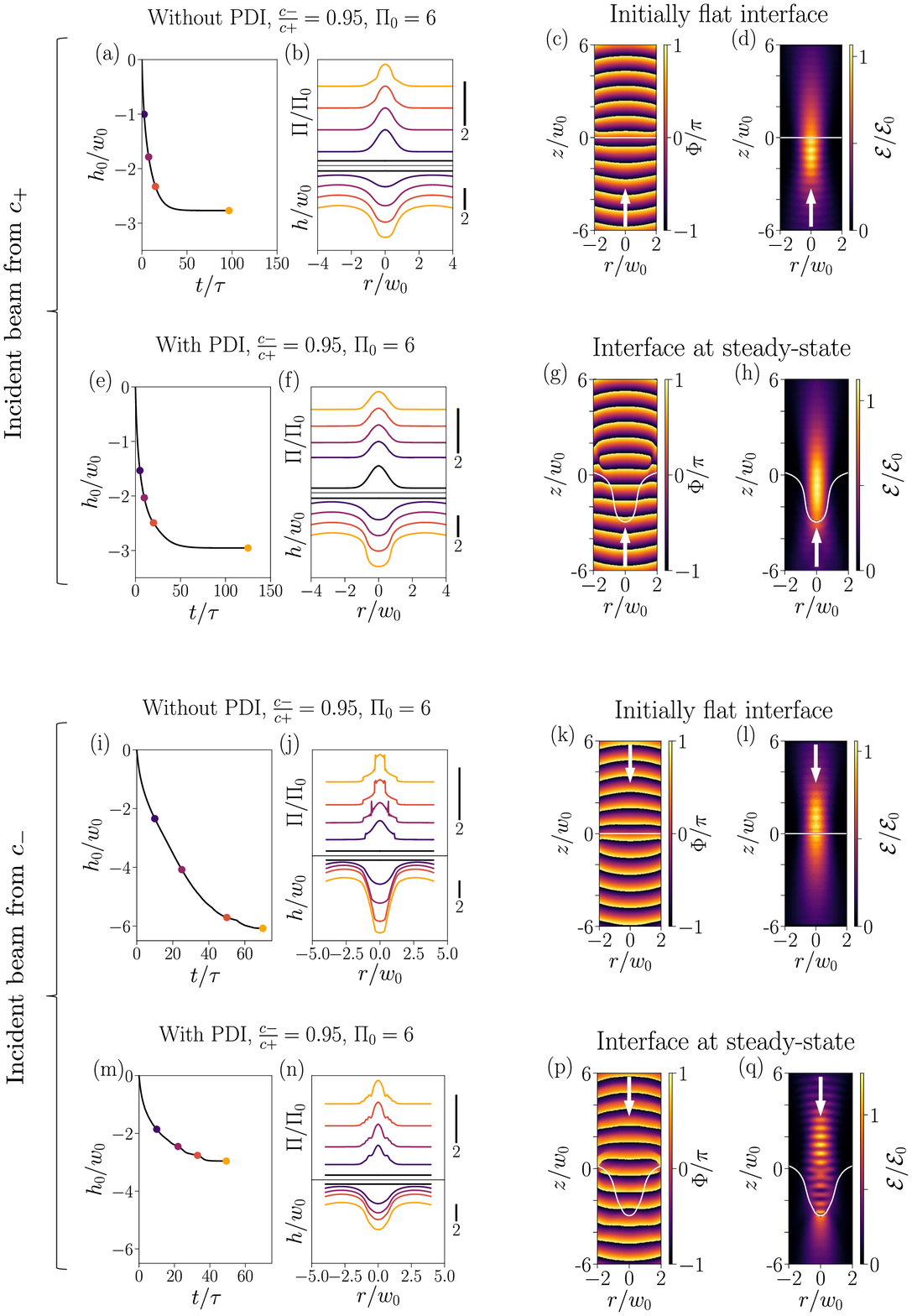}
\caption{\label{fig:dynamique2}
Computed morphogenesis of large interface deformation dynamics for celerity contrast $c_{-}/c_{+} = 0.95$, waveguiding parameter $V = 2.0$ with PDI (a,b, i, j) and without PDI (e)--(h), (m--q). (a)--(h) $c_{\text{from}} > c_{\text{to}}$, $\Pi_0 = 6$. (i)--(q) $c_{\text{from}} < c_{\text{to}}$, $\Pi_0 = 6$. Graphical representation is identical to that of Fig.~\ref{fig:dynamique1}.}
\end{figure*}

The comparison between \red{numerical} simulations with/without PDI confirms its paramount influence on both the shape and evolution of the irradiated interface. Noticeably, the steady shape of the irradiated interface shape shown in Fig.~\ref{fig:dynamique1}(h) qualitatively reproduces well the experimental observations on fluids such that $c_-/c_+ \simeq 0.5$, as shown in Fig.~\ref{fig:exp_morphologies}(a). 
Also we note in Fig.~\ref{fig:dynamique1}(h) the axial energy density modulation along the beam axis in fluid ``to''. It might be understood as the result of interferences between distinct guided modes (i.e., associated to distinct wavenumbers) propagating through the deformed interface acting as a multimodal waveguide. Such energy density modulation is also visible in Figs.~3(b) and 4(b) of Ref.~\cite{Bertin2012}, while the underlying multimodal propagation is demonstrated in Figs.~6(a), 6(b), 7(a) and 7(b) of Ref.~\cite{Bertin2012}.
%
%

\subsubsection{Large celerity contrast with $c_{\rm{from}} < c_{\rm{to}}$}

The evolution of the interface shape for $c_-/c_+=0.5$ and $c_{\text{from}} < c_{\text{to}}$, at $\Pi_{0} = 5$, is reported in the bottom-half part of Fig.~\ref{fig:dynamique1} in a similar manner as for the previously discussed case $c_{\text{from}} > c_{\text{to}}$.



In absence of PDI, the only qualitative difference between the case $c_{\rm{from}} < c_{\rm{to}}$ and the case $c_{\rm{from}} > c_{\rm{to}}$ lies in the emergence of a peak in the radiation pressure distribution along the deformed interface, as shown in Fig.~\ref{fig:dynamique1}(j). This peak corresponds to the coincidence of the local inclination of the interface with respect to the horizontal with the so-called total internal reflection (TIR) angle $\theta_{\rm TIR} = \arcsin(c_-/c_+)$~\cite{casner2003b}.

As shown in Fig.~\ref{fig:dynamique1}(m, n), when PDI is at work, a non-smooth evolution of the hump height is found as for the case $c_{\text{from}} > c_{\text{to}}$ and the interface deformation evolves toward a steady, slender and corrugated shape, also show in Fig.~\ref{fig:dynamique1}(q). This steady shape compares well with experimental observations on fluids such that $c_-/c_+ \simeq 0.5$ shown in Fig.\ref{fig:exp_morphologies}(b) .


%

\subsubsection{Small celerity contrast with $c_{\rm{from}} > c_{\rm{to}}$}

%
%

The morphogenesis of interface deformation for $c_-/c_+=0.95$ and $c_{\text{from}} > c_{\text{to}}$, at $\Pi_{0} = 6$, is reported in the top-half part of Fig.~\ref{fig:dynamique2} in a similar manner as for Fig.~\ref{fig:dynamique1}.



As expected, since the celerity contrast is small, the axial energy density modulations in fluid `from' are much less pronounced than in the case of large celerity contrast, as illustrated by the comparison between Fig.~\ref{fig:dynamique2}(d) and Fig.~\ref{fig:dynamique1}(d).
Although the interface shape evolution with and without PDI are qualitatively similar, the interface deformation is more cylindrical with PDI than without PDI, for which a bell-shaped deformation is observed.
This is attributed to the waveguiding effect of the deformation~\cite{Bertin2012}. The comparison between the steady shape visible in Fig.~\ref{fig:dynamique2}(h) and experimental observations on fluids such that $c_-/c_+ \simeq 0.9$ shown in Fig.~\ref{fig:exp_morphologies}(c), provides with a satisfying qualitative agreement.

%
%
\subsubsection{Small celerity contrast with $c_{\rm{from}} < c_{\rm{to}}$}

The evolution of the interface shape for $c_-/c_+=0.95$ and $c_{\text{from}} < c_{\text{to}}$, at $\Pi_{0} = 6$, is reported in the bottom-half part of Fig.~\ref{fig:dynamique2} in a similar manner as for the previously discussed case $c_{\text{from}} > c_{\text{to}}$.

\begin{figure}[t!]
    \includegraphics[width=1\columnwidth]{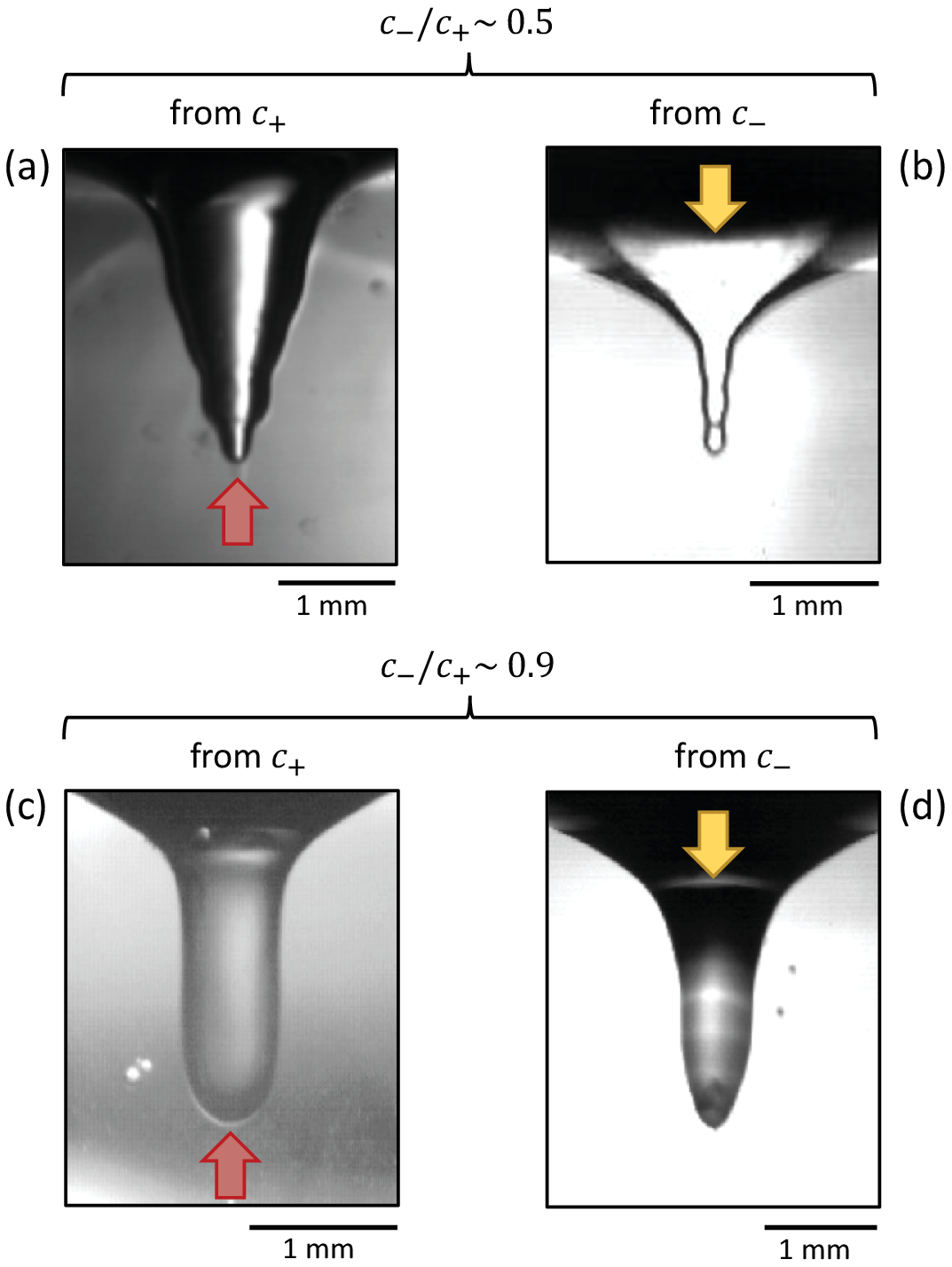}
    \caption{\label{fig:exp_morphologies} Experimentally observed morphologies at steady-state under acoustic irradiation at $2.25$~MHz frequency. (a) $20$~cSt Silicone oil / FC72 oil interface corresponding to $c_-/c_+=0.51$ and $V = 4.5$. 
    (b) $5$~cSt Silicone oil / $65~\text{wt}\%$ glycerol-water mixture interface corresponding to $c_-/c_+=0.54$ and $V = 4.5$. 
    (c) Water / kerosene interface corresponding to
    $c_-/c_+=0.88$ and $V = 2.6$. 
    (d) Water / kerosene interface corresponding to $c_-/c_+=0.88$ and $V = 2.6$. 
    }
\end{figure}

Although the hump height evolutions are qualitatively similar with and without PDI, the asymptotic regime is reached twice more rapidly with PDI than without and the asymptotic height is twice smaller with PDI than without, which substantially differs from the large contrast situation. Moreover, the steady deformation shapes noticeably differ depending on whether PDI is taken into account or not. Without PDI, the deformation tends to a cone, as shown in Fig.~\ref{fig:dynamique2}(j), whereas a bell-shaped deformation is found with PDI, as shown in Fig.~\ref{fig:dynamique2}(n, q).
These results demonstrate how strongly the PDI matters even when the celerity contrast is rather small. Also, a qualitative agreement is found between Fig.~\ref{fig:dynamique2}(q) and experimental observations on fluids such that $c_-/c_+ \simeq 0.9$ shown in Fig.~\ref{fig:exp_morphologies}(d).

\subsection{Slenderness transition}
%
\begin{figure}[b!]
    \includegraphics[width=\columnwidth]{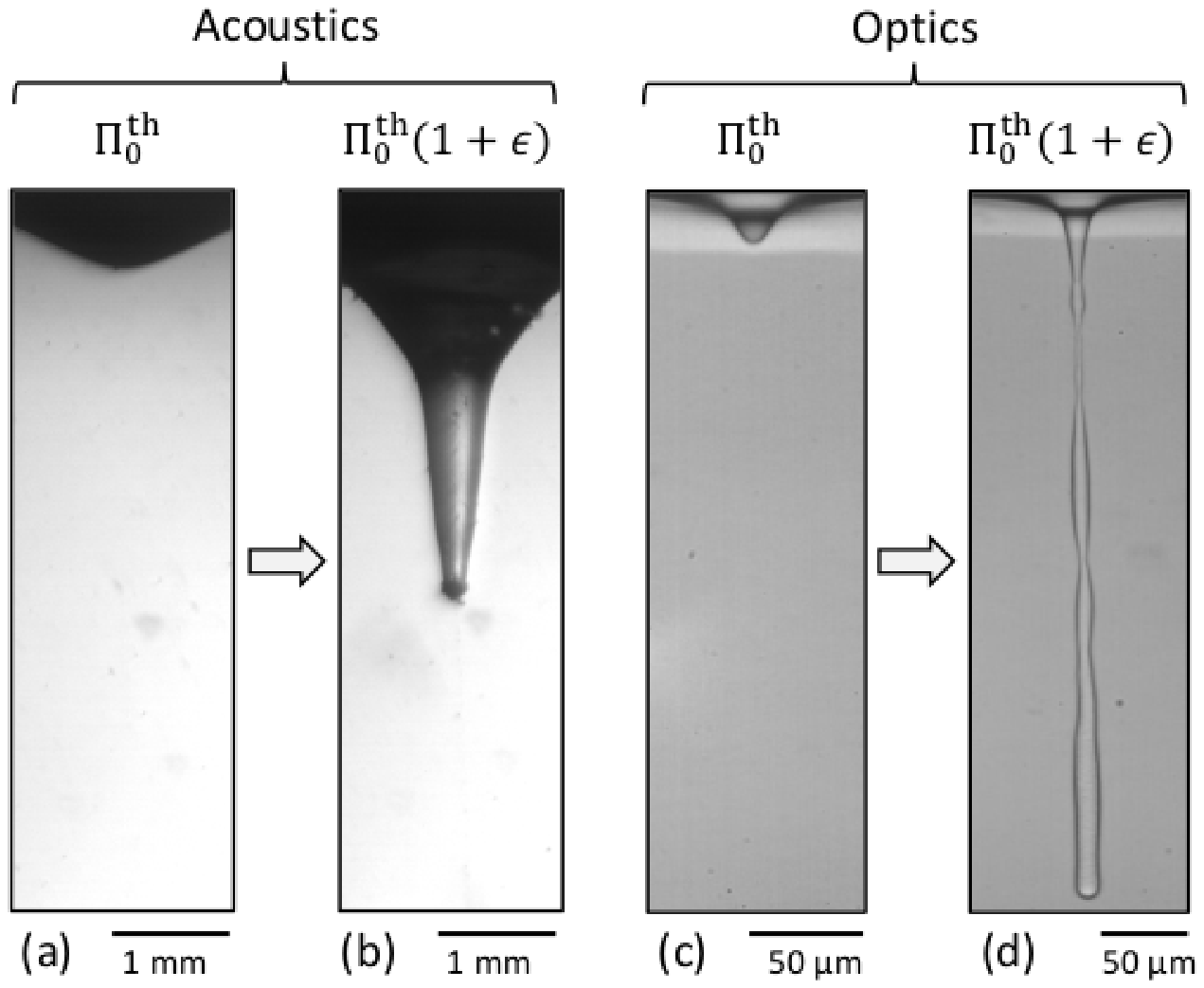}
    \caption{\label{fig:instability_exp} Experimental observation of the slenderness transition in acoustics at $2.25$~MHz frequency and in optics at $532$~nm wavelength. (a, b) Acoustic irradiation of a mineral oil / $32~\text{wt}\%$ glycerol-water mixture interface corresponding to $c_-/c_+ = 0.94$ and $V = 1.9$.
 (c,d) Optical irradiation of the interface between water-rich and oil-rich phases of a three-phase equilibrium (called as a Winsor phase) of a brine / AOT surfactant / n-heptane mixture~\cite{winsor1948,girot2019} corresponding to $c_-/c_+ = 0.96$ and $V = 34$. 
    Pictures of the interface deformations are shown at the onset of the transition threshold at $\Pi_0 = \Pi_0^{\text{th}}$ (a,c) and slightly above it at $\Pi_0 = \Pi_0^{\text{th}}(1+\epsilon)$ with \red{$\epsilon=0.02$} in panel (b) and $\epsilon=0.01$ in panel (d). In all panels the incident wave propagates from top to bottom ($c_{\rm from} < c_{\rm to}$).
    }
\end{figure}
%

\subsubsection{Background}

\begin{figure*}[t!]
    \includegraphics[width=2\columnwidth]{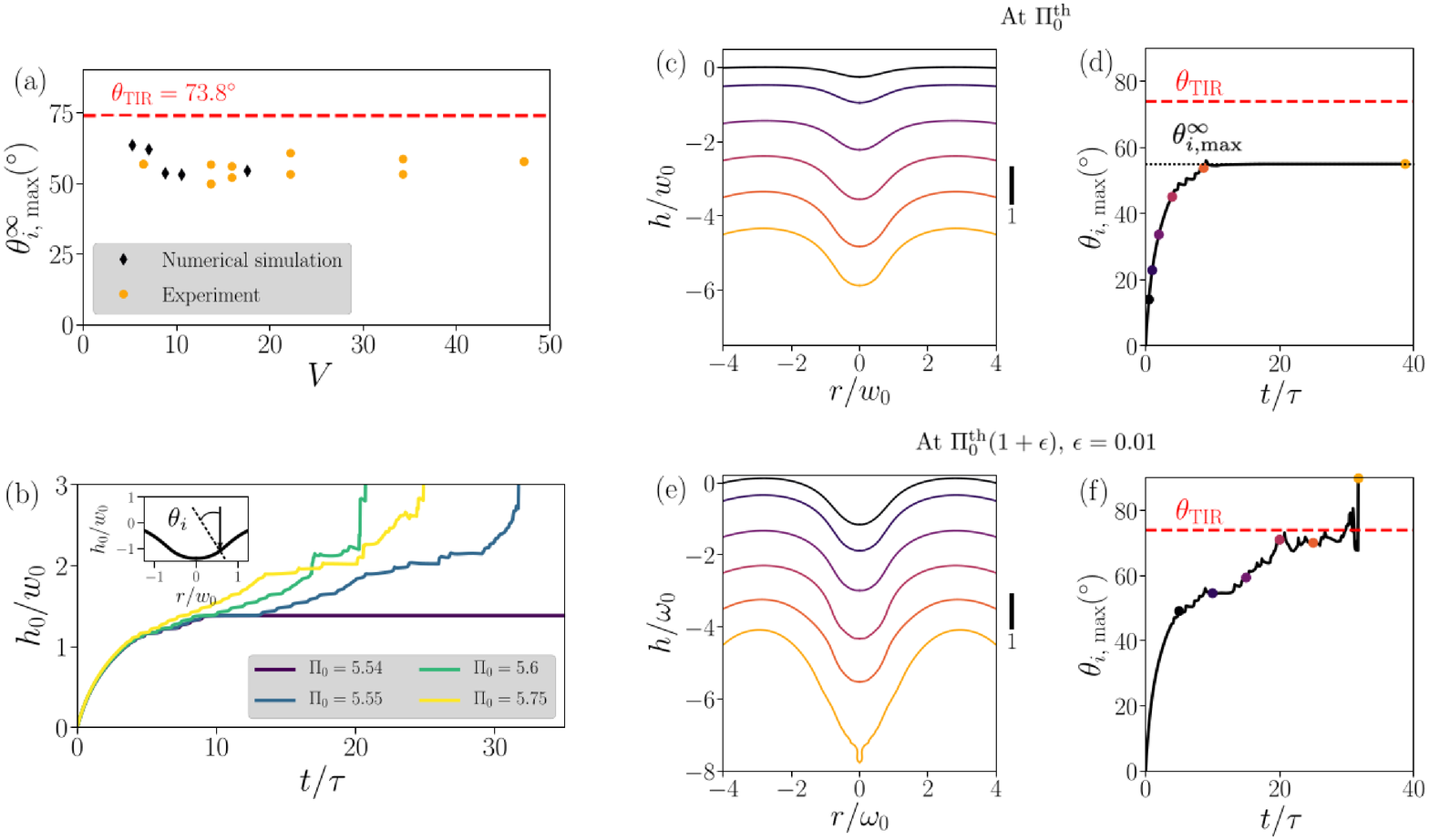}
    \caption{\label{fig:instability_simul}
    (a) Measured maximum value $\theta_{i,\text{max}}^{\infty}$ of the incidence angle $\theta_i$ of the rays on the steady deformed interface of a Winsor phase irradiated by a downward laser beam with $c_{\text{from}} < c_{\text{to}}$ at the threshold of the slenderness transition as function of the waveguiding parameter $V$ at constant value of the celerity contrast $c_-/c_+ = 0.96$. Orange circles: experimental values; Black lozenges: computed values; $\theta_{\text{TIR}}$: total internal reflection angle. (b) Computed time evolution of the dimensionless hump height for several values of $\Pi_0$ at and above the threshold of slenderness transition $\Pi_0^{\text{th}} = 5.54$, for $V=8.4$. Inset: sketch of the deformed interface illustrating the incidence angle $\theta_i$ of the rays on the deformed interface. (c,e) Deformation profiles at the threshold ($\Pi_0 = \Pi_0^{\text{th}}$) and just above ($\Pi_0 = 1.002~\Pi_0^{\text{th}}$) at $t/\tau = (0.5, 1, 2, 4, 8.7, 39)$ and $t/\tau = (5, 10, 15, 20, 25, 32)$, respectively. (d,f) Corresponding evolution of $\theta_{i,\text{max}}$. The colored markers refer to the instants at which the interface deformation profiles are shown in panels (c,e).}
\end{figure*}

Two decades ago, it was observed that an abrupt transition from moderately to strongly slender deformation occurs in the electromagnetic case only when $c_{\text{from}} < c_{\text{to}}$~\cite{Casner2003}. This so-called ``slenderness transition'' between two distinct regimes of deformation, was observed when a focused laser beam deforms an interface between the coexisting phases of a two-phase micro-emulsion close to its critical miscibility. A ray-optics based interpretation involving total internal reflection was suggested. Namely, it was proposed that the transition occurs when the inclination angle $\theta_i$ of the deformed interface with respect to the horizontal [see inset of Fig.~\ref{fig:instability_simul}~(b)] exceeds the angle of total internal reflection~\cite{Casner2003}, which is specific to the situation $c_{\text{from}} < c_{\text{to}}$.

Further investigations dedicated to the measurement of the maximal value for the steady-state incidence angle at the onset of the transition, $\theta_{i,\text{max}}^\infty$, showed that $\theta_{i,\text{max}}^\infty$ is unambiguously smaller than $\theta_{\text{TIR}}$~\cite{Wunenburger2006}, thus questioning the ray-optics approach. However, since the near-critical micro-emulsions involved in the first experimental observations were turbid~\cite{Casner2003}, both both radiation pressure and bulk radiation forces may contribute to the slenderness transition. Since then, bulk radiation forces associated with turbidity have been shown to induce a jet transition (i.e., the sudden occurrence of droplet emission at the tip of a needle-like deformation~\cite{wunenburger2011}) which shares several common features with the slenderness transition~\cite{chraibi_excitation_2013,Chesneau2020}. The question of whether the slenderness transition can be triggered solely by radiation pressure is still open.

Here we provide new insight into the slenderness transition mechanism by combining \red{novel} experiments and numerical investigation in the absence of bulk radiation forces, within a framework encompassing both for acoustic and electromagnetic waves.

\subsubsection{Experimental observations}

Two experimental observations of slenderness transition respectively driven by acoustic and optical beams are reported in Fig.~\ref{fig:instability_exp} at the onset of the transition threshold, which corresponds to $\Pi_0 = \Pi_0^{\rm th}$, and slightly above the threshold with $\Pi_0 = \Pi_0^{\text{th}}(1+\epsilon)$, where $\epsilon$ is a small parameter. In both cases the celerity contrast is small and almost identical, namely, $c_-/c_+ = 0.94$ in acoustics and $c_-/c_+ = 0.96$ in optics, hence providing a playground for a generic comparison between experiments and numerics\red{, which calls for two remarks. First, in acoustics, the thermoviscous dissipation prevents us from achieving an experiment virtually free of bulk radiation forces (i.e. free of acoustic streaming) and, in our experiments, the waveguiding parameter $V$ can only be tuned by changing the fluids. Second, in optics it is possible to deal with fluids virtually free from bulk forces~\cite{girot2019} and in our experiments $V$ can be readily tuned by changing the beam waist $w_0$. Therefore, because we aim to test the impact of PDI only on the slenderness transition, hereafter we focus on the electromagnetic case and we present an experimental parametric study that is compared to numerical simulations.}

The optical experiment is carried out using visible radiation on a two-phase fluid composed a water-rich phase coexisting with an oil-rich phase of a three-phase equilibrium of a brine / dioctyl sodium sulfosuccinate (AOT) surfactant / n-heptane mixture~\cite{winsor1948} called a Winsor phase. Such a fluid interface is characterized by (i) a small interfacial tension, allowing the slenderness transition to be triggered using a focused beam emitted by a tabletop continuous-wave laser, and (ii) an optical transparency of both phases in contact, which ensures that bulk forces can be safely discarded.

The experiments consists in irradiating the fluid using a Gaussian laser beam focused on the fluid interface at rest and in observing the steady deformed interface. The threshold power value $P^{\text{th}}$ at which the slenderness transition occurs is then determined by dichotomy with a $1~\%$ accuracy in a wide range of values of $w_0$ at fixed celerity contrast $c_{-}/c_{+}=0.96$ with $c_{\rm from} < c_{\rm to}$. The angle $\theta_{i \text{max}}^{\infty}$ is then evaluated at the inflection point of the steady interface deformation by image processing. The variation of $\theta_{i \text{max}}^{\infty}$ versus $V$ is displayed in Fig.~\ref{fig:instability_simul}(a). We find that $\theta_{i,\text{max}}^{\infty}$ is nearly independent of $V$ and that $\langle \theta_{i,\text{max}}^{\infty}  \rangle = 55.3^\circ$ (with $3.3^\circ$ standard deviation) is smaller than $\theta_{\text{TIR}} = 73.8^\circ$ by more than $18^{\circ}$. This demonstrates experimentally that the slenderness transition is not driven by the total internal reflection phenomenon and does not require bulk radiation forces. 
%
\subsubsection{Numerical investigation}
The numerical investigation of the slenderness transition, which is only observed with PDI and with $c_{\rm from} < c_{\rm to}$, is conducted by mimicking the experiments presented in the previous section. Accordingly, we determine the threshold dimensionless power value $\Pi_{0}^{\rm th}$ at which the slenderness transition takes place. A typical numerical experiment is illustrated in Fig.~\ref{fig:instability_simul}(b) where the evolution of the on-axis height of the deformed interface is shown for different values of $\Pi_{0}$ at $c_-/c_+=0.96$ and $V=8.4$. The sudden occurrence of a diverging behavior of the deformation height as $\Pi_{0}$ increases from 5.54 to 5.55 illustrates the abrupt nature of the slenderness transition with respect to the control parameter $\Pi_{0}$ and justifies the $0.2~\%$ accuracy of the determination of $\Pi_{0}^{\rm th}$. A detailed picture of the interface evolution at the threshold and slightly above is shown in Figs.~\ref{fig:instability_simul}(c-f). The dimensionless deformation profiles $h(r)/w_0$ vs $r/w_0$ are displayed in panels (c) and (e) at several dimensionless instants. Also the corresponding evolution of the interface inclination angle $\theta_{i,{\rm max}}$ at the inflexion point are shown in panels (d) and (f). For $\Pi_0 = \Pi_{0}^{\text{th}}$, $\theta_{i,\text{max}}$ increases, then saturates at a value $\theta_{i,\text{max}}^\infty$ significantly smaller than $\theta_{\text{TIR}}$, while for $\Pi_{0} = 1.002 \, \Pi_{0}^{\text{th}}$, the increase of $\theta_{i, \text{max}}$ slows down when approaching $\theta_{i,\text{max}}^{\infty}$, then revives up to $\theta_{\text{TIR}}$. Finally, a thin tip grows at the deformation end, as shown in Figure~\ref{fig:instability_simul}(e), leading to  $\theta_{i,\text{max}} \simeq 90^{\circ}$, as shown in Figure~\ref{fig:instability_simul}(f).


\begin{figure}[t!]
    \includegraphics[width=0.9\columnwidth]{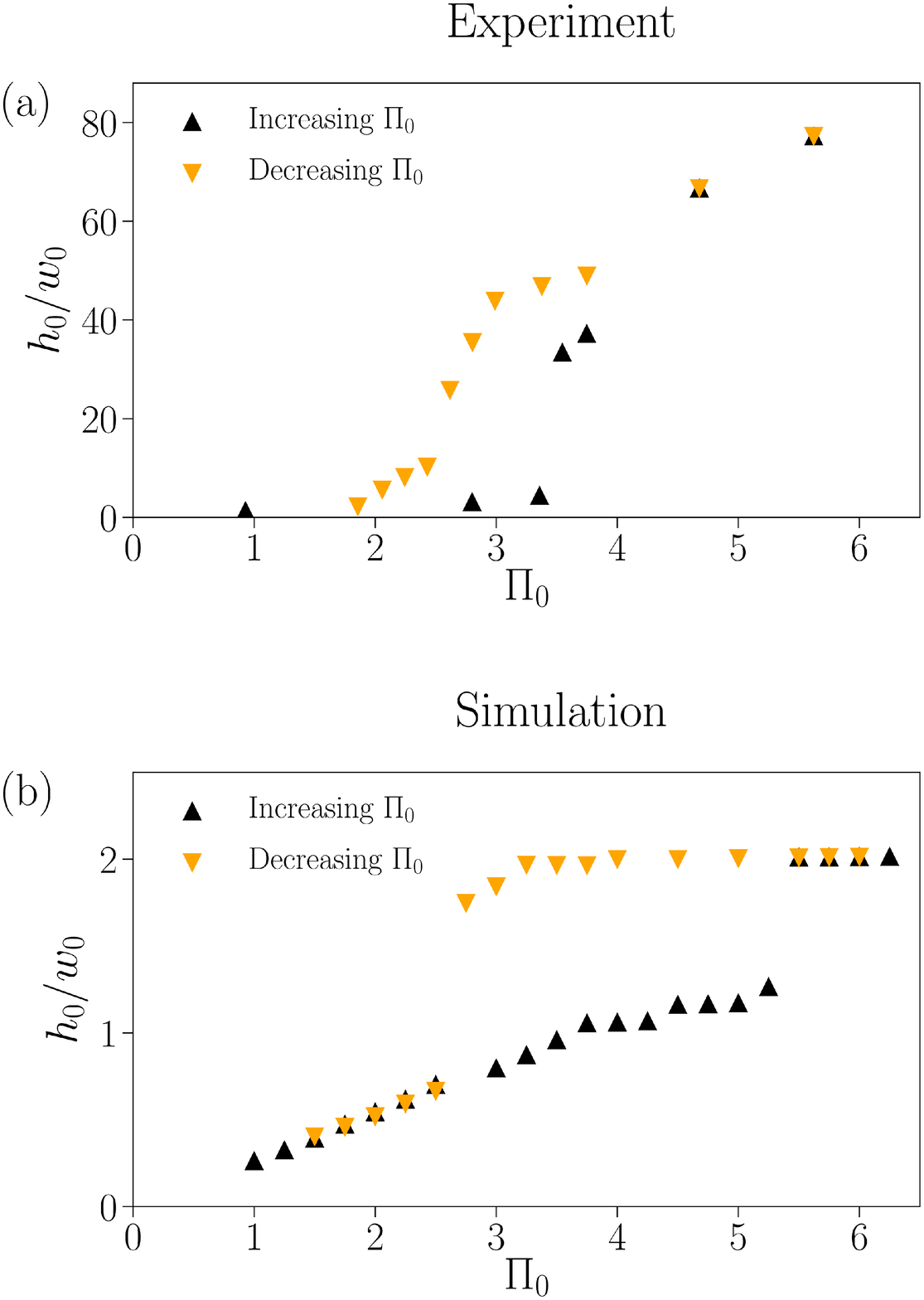}
    \caption{
    \label{fig:hysteresis}
    (a) Experimentally observed hysteretic behavior of the deformation height $\tilde{h}$ versus $\Pi_0$ of the interface of a Winsor phase such that $c_-/c_+= 0.96$, $V=13.7$. 
    (b) Computed hysteretic behavior for $c_-/c_+= 0.93$, $V=6.9$.
    }
\end{figure}

Since $V$ is varied by varying $k w_0$ at fixed $c_-/c_+$, increasing $V$ requires to increase the spatial resolution of the simulations. Given the associated computational cost, our investigations are restricted to the typical range $2 < V < 18$. The numerically computed dependence of $\theta_{i,\text{max}}^{\infty}$ versus $V$ is displayed in Fig.~\ref{fig:instability_simul}(a), which quantitatively agrees with the experimental measurements. This confirms that the slenderness transition does not require bulk radiation forces\red{, cannot be explained by the sole ray-optics approach, and is actually driven by PDI. In addition, Figs.~5(a) and 5(b) also support preceding conclusions in the acoustic domain. Indeed, for this pair of fluids, $\theta_{\rm TIR} \simeq 70^\circ$ whereas the transition occurs at $\theta^{\infty}_{i,{\rm max}} \simeq 30^\circ < \theta_{\rm TIR}$.}
\subsection{Hysteresis of the slenderness transition}
%
Beyond the slenderness transition, by cycling the beam power up and down in the typical range $0 < \Pi_{0}/\Pi_{0}^{\text{th}} < 2$, we evidenced a hysteretic behavior of the deformation height versus $\Pi_0$, as shown in Fig.~\ref{fig:hysteresis}(a). Experiments correspond to $V=13.7$ and the hysteresis phenomenon is also observed for larger values of $V$. The relative width of the hysteresis cycle is $\sim 50~\%$ and its lower and upper branches are characterized by a height ratio $\sim 10$. 
Noticing that a hysteresis is a generic manifestation of the feedback action of a system on its driving parameter, these experimental results tend to further highlight the role of PDI that can be also explored numerically. For this purpose, we simulate the interface response to a cycling of the dimensionless power in a configuration as close as possible to the experimental conditions. \red{Recalling that the computational limitations prevent us from assessing deformation heights as large as those observed experimentally ($h_0/w_0 = 10-100$), we chose material parameters enabling the demonstration of the existence of an hysteretic behavior, namely $c_-/c_+=0.93$ and $V = 6.9$. The results are shown in Fig.\ref{fig:hysteresis}(b). A hysteresis cycle is numerically obtained,
which demonstrates that PDI is required for describing a hysteresic behavior.}

\section{Concluding remarks}

As illustrated by the Eqs.~\eqref{eq:ADac4} and~\eqref{eq:ADem4}, the interface deformations induced by the radiation pressure of an acoustic or an electromagnetic beam can be described in a universal manner using 4 parameters (in the limit of small density contrast regarding acoustic fields), which makes their thorough numerical investigation rather challenging. For this reason, in this work we have numerically investigated a few phenomena such as the deformation morphologies, the slenderness transition and its hysteresis, using parameter values as close as possible to those encountered in experiment. In overall, we found qualitative agreement between numerical simulations and experiments only when the propagation-deformation interplay is taken into account, hence pinpointing its crucial role in the interface deformation morphogenesis \red{in situations where streaming flows induced by bulk forces---which can occur both in acoustics and optics---can be neglected}. Therefore, the numerical model presented here constitutes a valuable tool for further investigations of these phenomena.

\bibliographystyle{apsrev}


\end{document}